\documentclass[doublecol]{epl2} 

\usepackage{graphicx}
\usepackage{float}
\usepackage{amsmath, amssymb}
\usepackage{lmodern}
\usepackage[T1]{fontenc}
\usepackage[utf8]{inputenc}
\usepackage[english]{babel}
\usepackage{color}

\title{Is the public goods game a chaotic system?}
\shorttitle{Is the public goods game a chaotic system?}

\author{D. Bazeia\inst{1}, M. J. B. Ferreira\inst{2}, and B. F. de Oliveira\inst{3}}
\shortauthor{Bazeia, Ferreira, de Oliveira}

\institute{
\inst{1}Departamento de Física,     Universidade Federal da Paraíba,  58051-970 João Pessoa, PB, Brazil\\
\inst{3}Departamento de Matemática, Universidade Estadual de Maringá, 87020-900 Maringá,     PR, Brazil\\
\inst{3}Departamento de Física,     Universidade Estadual de Maringá, 87020-900 Maringá,     PR, Brazil\\
}

\pacs{87.23.Kg}{Dynamics of evolution}
\pacs{87.23.Cc}{Population dynamics and ecological pattern formation}
\pacs{05.45.-a}{Nonlinear dynamics and chaos}

\abstract{This work deals with the time evolution of the Hamming distance density for the public goods game. We consider distinct possibilities for this game, which are exactly described by a function called $q$-exponential, {\color{black} that represents a deformation of the usual exponential function parametrized by $q$, suggesting that the system belongs to the class of weakly-chaotic systems when $q < 1$}. These possibilities are related to the amount of players allowed in each game.}

\begin{document}

\maketitle

\section{Introduction}

Biodiversity is a complex and very hot topic in the present days. Rules that guide evolution and stability of systems based on biodiversity have received important contributions in the last 20 years \cite{ref-04,ref-05}. In particular, the investigations \cite{2002-Kerr-N-418-171, 2004-Kirkup-Nature-428-412, 2007-Reichenbach-N-488-1046} have shown the importance of the rock-paper-scissors (RPS) rules to control and maintain simple models of biodiversity in dynamical evolution. 

In a more recent work, in Ref. \cite{2017-Bazeia-SR-7-44900} some of us have studied the presence of chaos in the standard three species model under the rules of the RPS game, through the use of the Hamming distance concept \cite{1950-Hamming-BSTJ-29-147}. This study was also explored in \cite{2017-Bazeia-EPL-119-58003} in several distinct scenarios, where the number of species was increased from three to four, five, six, seven, eight, nine and ten species. An interesting result is that as the system evolves in time, the profile of the Hamming distance density seems to show an universal behavior, increasing until reaching an asymptotic value that covers a significant portion of the system, usually, larger than 65 percent of its entire contents. {\color{black} We also want to emphasise that in Ref. \cite{2007-Reichenbach-N-488-1046}, the authors have shown that in the RPS model, when one increases the rule of mobility to very high values, the system breaks biodiversity. Inspired by this, in \cite{2017-Bazeia-EPL-119-58003} the investigation has also shown how the Hamming distance behaves for very high values of mobility. The results confirmed that when the rule of mobility is increased to higher and higher values, the Hamming distance trivializes, going to zero or unity, as expected. Of course, in this case the system engenders no chaotic behavior anymore. One also notices that the Hamming distance was further investigated in Ref. \cite{Mugnaine_2019} under similar questioning, giving the same qualitative results.}

The success of the use of the Hamming distance to uncover the presence of chaos in the dynamical evolution of models based on the RPS rules has motivated us to further explore the issue in other similar systems. Among distinct possibilities, we have encountered the challenging option to consider the so called public goods game (PGG), which was studied under distinct motivation by several authors, in particular, in \cite{Hauert2002,Santos2008, 2009-Szolnoki-PRE-80-056109,2011-Szolnoki-PRE-84-046106,PhysRevE.84.047102,ref-01,ref-02,ref-03}, and in references therein. The game considers many players, which belong to two different kinds, either cooperators or defectors, possessing opposite strategies. {\color{black} It is important to state that some models consider strategies other than cooperators or defectors, as in reference \cite{Hauert2002}}. Cooperators are those who contribute an amount $1$ to a common pool shared by other players, while defectors contribute nothing. Each round of this game is played by a group $\mathcal{G}$ of players, being $G$ the size of such a group.

The total amount collected at each round is multiplied by an enhancement factor $r$ and then equally divided among all the players within $\mathcal{G}$. In other words, all the players receive a payoff regardless of their strategy. It is a known fact from the literature that after many rounds and for sufficient high enhancement factor $r$ the system may succeed and eventually all the players become cooperators. Likewise, if the enhancement factor is too low, the system declines and eventually all the players become defectors, leading to the effect known as the tragedy of the commons.  For intermediate values of $r$, cooperators and defectors may coexist on a stable phase where the proportion of each kind depends on $r$. This quantitative behavior depends on the size of the group $G$ as pointed out by \cite{PhysRevE.84.047102}. In this work we shall investigate the chaotic behavior of this game and how it is affected by the group size $G$. The chaotic behavior is quantified by monitoring the Hamming distance of two identical systems evolving from slightly different initial condition. The Hamming distance in this case refers to the counting of players mismatching strategies on the two systems. 

To implement the investigation, we organize the work as follows. In section {\bf Model} we describe the classical spatial PGG model choosing a specific two-dimensional square lattice and the dynamics of such a model and how it is affected by the group size. We go on and in the section {\bf Hamming Distance} we describe the Hamming distance, highlighting the main characteristics as an important tool to be developed in this work. The most relevant results are then presented in section {\bf Results}, and we close this study in the section {\bf Conclusion}. 

\section{Model}

In this section we define the PGG model as  follows. 

\subsection{The Lattice}
The public goods game can be modeled by a stochastic simulation where the players take place on the sites of a two-dimensional square lattice sized $L \times L$ with periodic boundary conditions. At each site $c$ is associated a number $s(c)=1$ for cooperators or $s(c)=0$ for defectors. 

\begin{figure}[!htb]
	\centering
	\includegraphics[scale=1.2]{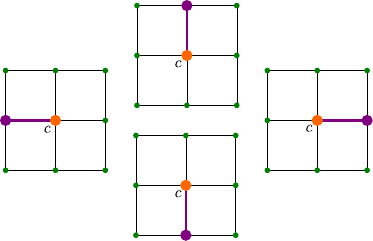}
	\caption{On a square lattice, every player $c$ (orange dot) has exactly four nearest neighbors. Here one can visualize all the $4$ rounds $c$ participates in for $G=2$. The purple dots represents the other players in the group.}
	\label{fig1}
\end{figure}

The number of players participating of every round, i. e., the size of the group, is specified by the parameter $G$. On a square lattice every player $c$ has exactly four nearest neighbors. For $G=2$, for example, every player $c$ participates in four different rounds, as it can be easily seen in fig. \ref{fig1}. 

\begin{figure}[!htb]
	\centering
	\includegraphics[scale=1.2]{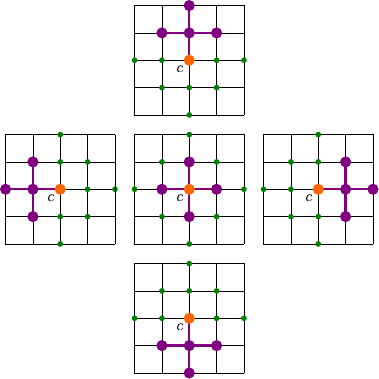}
	\caption{All the $5$ rounds $c$ participates in for $G=5$. The purple dots represents the other players in the group.}
	\label{fig2}
\end{figure}

Likewise, for $G=5$ it can be seem from fig. \ref{fig2} that every player $c$ participates in $5$ rounds.

\begin{figure}[!htb]
	\centering
	\includegraphics[scale=1.2]{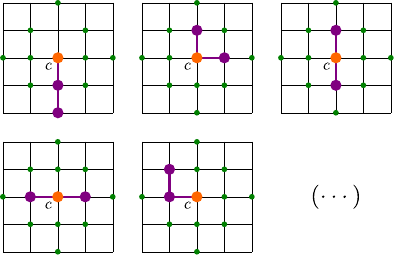}
	\caption{Some examples of rounds $c$ participates in for $G=3$ (from a total amount of $18$ rounds). The purple dots represents the other players in the group.}
	\label{fig3}
\end{figure}

{\color{black} There are other possibilities, for example, $G=3$ and $G=4$. In these cases, the combinatorial approach has to be taken into account. For $G=3$, the total amount of rounds is $18$, and for $G=4$, it is $16$. In fig. \ref{fig3} one can visualize a few examples of these rounds for $G=3$. The case $G=4$ is similar.}

\subsection{The Dynamics}

The dynamics of this model is defined via stochastic simulations where at each Monte Carlo Step (MCS) a random player $c$ is selected, which we call the central player. The player $c$ participates of all the rounds and after each round the player $c$ will receive a payoff
\begin{equation*}
	p(c) = \frac{r}{G} \sum_{x\in \mathcal{G}}s(x) - s(c)\;. 
\end{equation*}
The total payoff of $c$ will be the sum of all the payoffs collected by $c$ from all the rounds it participates in. After each game, a player $c$ may decide to adopt the strategy of a randomly chosen neighbor $n$ with a probability $W_{c\to n}$, given by the Fermi-Dirac distribution
\begin{equation}
	W_{c\to n} = \dfrac{1}{1+\exp\left[(p(c)-p(n))/K\right]}\ ,
	\label{eq1}
\end{equation}
where $K$ represents an irrationality coefficient. It is used to include uncertainty on the strategy; see Ref. \cite{2007-Szabo-PR-446-97}. In brief, if $K$ is very small, player succeeds in enforcing its strategy, but as $K$ increases, strategies performing worse may also be adopted.

\begin{figure}[!htb]
	\centering
	\includegraphics{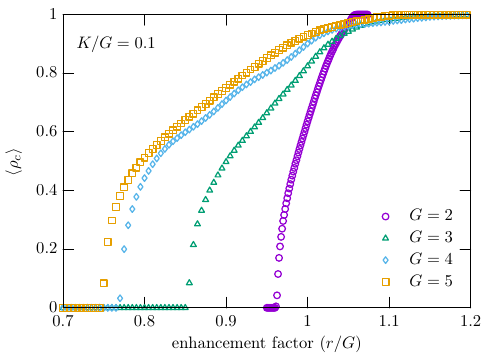}
	\caption{Average density of cooperators over 5000 simulations after 5000 generations as a function of $r/G$. These results are similar to the ones found in \cite{2009-Szolnoki-PRE-80-056109}. Here we have set $K/G=0.1$ and $L=500$.}
	\label{fig4}
\end{figure}

In order to implement the numerical simulations, we have introduced generation, which is defined as the time associated to the number of MCS (on average) necessary to all players to be selected at least once, which in this case is $L^2$. Moreover, in order to quantify the behavior of such a model as a function of the parameter $G$ we performed $5000$ simulations for each of these cases. The results obtained are shown in fig. \ref{fig4} and they are in agreement to the ones found in \cite{2009-Szolnoki-PRE-80-056109}.
In this figure, one shows the average density of cooperators $\langle\rho_c\rangle$ after $5000$ generations as a function of $r/G$ while keeping the ratio $K/G=0.1$ fixed. Note that the smaller the group is the higher the enhancement factor $r$ has to be in order to avoid the tragedy of the commons. Smaller groups also require higher enhancement factors to reach the equilibrium phase where all the players become cooperators. That means smaller groups are more likely to decline leading to the tragedy of the commons. As it can be noticed from fig. \ref{fig4}, the range of values of $r$ allowing the system to reach a coexistence phase is shorter for smaller groups as well.

\begin{figure}[!htb]
	\centering
	\includegraphics{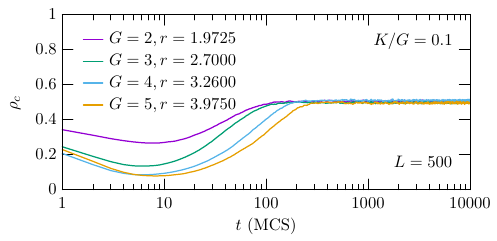}
	\caption{Density of cooperators as a function of the time for over
	$10^4$ generations. Note the behavior are quite different at the
	beginning but after 300 generations the differences and fluctuations
	become very small. We set $r=1.9725$ for $G=2$, $r=2.7000$ for $G=3$, $r=3.2600$ for $G=5$ and $r=3.9750$ for
	$G=5$. These choices were made in order to get $\langle \rho_c \rangle$ the closest to $0.5$.}
	\label{fig5}
\end{figure}

Another result in depicted in fig. \ref{fig5}, in which one shows the evolution of the density of cooperators for different values of $G$ for over $10^4$ generations. For these simulations the enhancement factors were chosen in each case such that the asymptotic values are the closest to $0.5$ as it will be clear in the following (see the ending paragraph). At the beginning the distinct cases exhibit different behavior until nearly $300$ generations. From this point on their differences and fluctuations become very small.

\begin{figure}[!htb]
	\centering
	\includegraphics{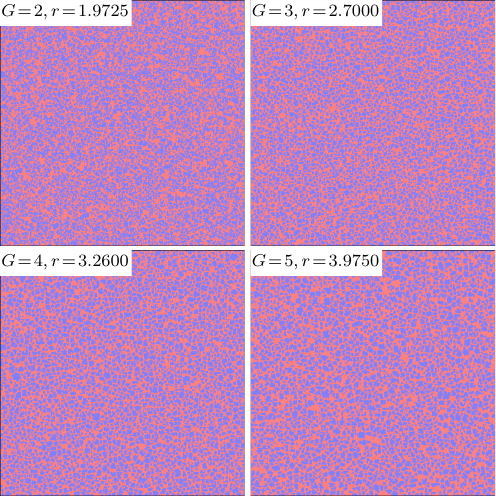}
	\caption{Snapshots displayed after $5000$ generations for $G=2$ ($r=1.9725$), $G=3$ ($r=2.7000$), $G=4$ ($r=3.2600$) and $G=5$ ($r=3.9750$) with $L=500$. Here blue colour represents cooperators while red colour represents defectors.}
	\label{fig6}
\end{figure}

In the fig. \ref{fig6} one can see snapshots of simulations obtained after $5000$ generation for different values of $G$. 

We are now ready to quantify the chaotic behavior of such a model for different values of the parameter $G$. As suggested in \cite{2017-Bazeia-SR-7-44900}, here we also study the Hamming distance in order to measure the distance between two almost identical system with slightly different initial conditions, as we discuss in the next Section.

\section{Hamming Distance}

In order to analyze the chaotic behavior of such models we generate a system on a square lattice $L\times L$ starting with a random initial condition. The system then evolve for over $5000$ generations in order to reach a stable phase, as shown in fig. \ref{fig5}. This new configuration is the starting point for our simulation. A copy of the system is made, but one single player of such a copy is randomly chosen to have its strategy flipped. In this way the simulation starts from two almost identical system, with only one single player mismatching strategy. The two systems then evolve following the same stochastic rules. The Hamming distance $\Delta H(t)$ is used to quantify the distance of the two systems as a function of the time. We define the Hamming distance $\Delta H(t)$ by the number of mismatching within the two systems. Note that initially the Hamming distance is one, $\Delta H(0)=1$. 

{\color{black}Before we proceed, since the Hamming distance is important tool for the present study, let us explain the concept more carefully. In Ref. \cite{1950-Hamming-BSTJ-29-147}, Hamming suggested a simple way to distinguish quantities such as vectors,
matrices, etc. This is usually called the Hamming distance, and we can illustrate the concept with binary vectors, for instance. It simply counts the number of sites in the second vector that do not match with the corresponding sites of first one. In the example with the two vectors (0,1,0,1,1) and (0,0,1,1,0), the Hamming distance is three. In the present study, we use the Hamming distance to measure the difference between two systems in the square lattice, seeing them as two square matrices $L\times L$ and counting the number of sites that are different. This procedure was used before in Refs.\cite{2017-Bazeia-SR-7-44900,2017-Bazeia-EPL-119-58003}, and here we follow the same strategy.}

Usually, chaotic systems are quantified by Lyapunov coefficient $\lambda_1$, defined by
\begin{equation}
	\Delta H(t) \sim \Delta H(0) \exp(\lambda_1 t)\ ,
	\label{eq2}
\end{equation}
for $\Delta H(0) \rightarrow 0$ and $t \rightarrow \infty$. However, some systems are called weakly-chaotic (in the sense that the rate of separation of infinitesimally close trajectories in phase space is not exponential)\cite{cite-key,1997-Tsallis-CSF-8-885}, and this happens when $\Delta H(t)$ is given by a $q$-exponential instead
\begin{equation}
	\Delta H(t) \sim \Delta H(0) [1 + (1-q)\lambda_q t]^{1/(1-q)}\ ,
	\label{eq3}
\end{equation}
where $q \in \mathbb{R}$. Note that in the limit $q\to 1$ equation \eqref{eq2} is recovered.

For the purposes of this work it is more convenient to analyze the stable phases where cooperators and defectors coexist in the same amount, i. e., in each case we set $r$ such that $\langle \rho_c\rangle$ is the closest to $0.5$. For this reason, and based on the data displayed in fig. \ref{fig4}, we have chosen $r=1.9725$ for $G=2$, $r=2.7000$ for $G=3$, $r=3.2600$ for $G=4$ and $r=3.9750$ for $G=5$.

\section{Results}

\begin{figure}[!htb]
	\centering
	\includegraphics{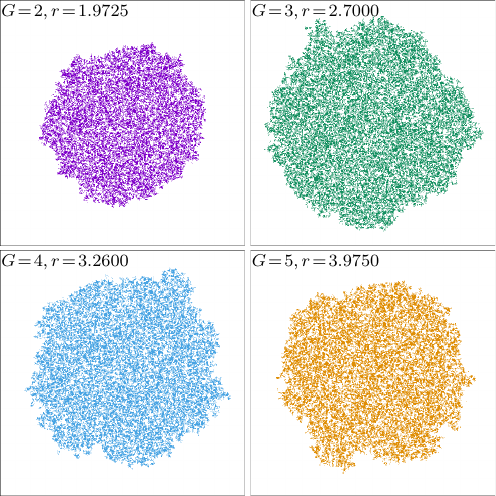}
	\caption{Snapshots showing the mismatches after 700 generations. Here $L=500$ and the enhancement factors are indicated in the
	pictures. The mismatches represents $18\%$ of the lattice for $G=2$, $26.4\%$ for $G=3$, $23.3\%$ for $G=4$ and $24\%$ for $G=5$.}
	\label{fig7}
\end{figure}

As the systems evolve the mismatches between the two systems spread over the lattice as shown in fig. \ref{fig7} for different values of $G$. In this picture the colored dots represent the mismatches for a simulation after 700 generations. The number of mismatches represents $18\%$ of the lattice in the case where $G=2$, and $26.4\%$ for $G=3$, and $23.3\%$ for $G=4$ and $24\%$ of the lattice for $G=5$, indicating that this number grows faster for groups with higher number of rounds, as it is the case for $G=3$.

\begin{figure}[!htb]
	\centering
	\includegraphics{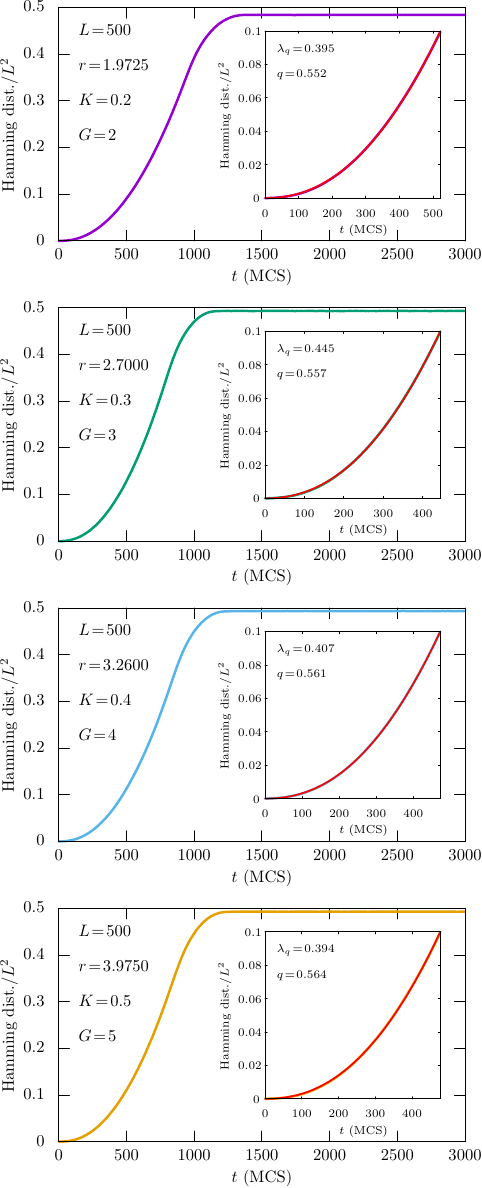}
	\caption{Hamming distance per $L^2$ for different values of $G$ as a function of the time. These pictures were obtained by averaging the Hamming distance for over 500 simulations. The inserts show the fit of the $q$-exponential for the first 500 generations. The parameters used in these simulations are also shown in the figure.}
	\label{fig8}
\end{figure}

The Hamming distance normalized by $L^2$ as a function of the time is shown for different values of $G$ in fig. \ref{fig8}.
These pictures represent the Hamming distance averaged over $500$ simulations each. It is clear in all cases that the Hamming distance grows slower at the beginning and it gets faster up to $1000$ generations, after that reaching an asymptotic value close to $50\%$. Evidently these functions cannot be properly fitted by an ordinary exponential function, meaning these models do not accomplish a strict chaotic behavior. However, it suggests a weak-chaotic behavior, which can be fitted by a generalized $q$-exponential instead.

As one can see in all the cases, the $q$-exponential fits perfectly the data in all cases. The inserts in fig. \ref{fig8} show these fits for the first $500$ generations. 

These fits were obtained by the following procedure: first we verify how many generations it takes in order to the Hamming distance per $L^2$ be equal to $0.1$ and then we fit the data using equation \eqref{eq3} with $\Delta H(0)=1$. The value $0.1$ was chosen in order to avoid boundary effects. Note that in both cases $q < 1$ and $\lambda_q > 0$, meaning that the PGG systems are weakly sensitive to the initial conditions.

{\color{black} The Lyapunov coefficient is usually considered to measure the sensitivity of the system to the initial conditions. In this sense, the greater positive Lyapunov coefficient, the more sensitive the system is to the initial conditions}. However, for some systems the measurement of this coefficient can be quite hard, so here we decided to use the approach described in \cite{2017-Bazeia-SR-7-44900,2017-Bazeia-EPL-119-58003}, using the Hamming distance concept. This procedure can be used to separate deterministic and stochastic behavior \cite{2017-Bazeia-SR-7-44900}, allowing to obtain the Lyapunov coefficient.

The topology of the lattice constrains the spreading of the mismatchings over the lattice. For this reason we are using a lattice with simple topology, and the $q$-exponential along with the Hamming distance to achieve the Lyapunov coefficient. According to \cite{1997-Tsallis-CSF-8-885}, for $q=1$, for example, the system is called chaotic in the standard sense while for $q<1$ the system is called weakly-chaotic.

Even though the Lyapunov coefficients were computed and displayed in fig. \ref{fig8}, it is not possible to compare these systems via Lyapunov coefficient since the functions used to fit the Hamming distance are different, having distinct values of $q$. As it is shown in fig. \ref{fig8}, the parameter $q$ increases as we increase the group size, $G$.

We can notice from fig. \ref{fig8} that the Hamming distance increases faster for systems with a higher number of rounds, as it is the case for $G=3$, with 18 rounds, in this case the mismatches reaches $10\%$ of the lattice after 445 generation. The next in the sequence is the case with $G=4$, with 16 rounds, which reaches the same $10\%$ of the lattice after 473 generations. After that, for $G=5$, with 5 rounds, it reaches $10\%$ after 477 generations, finally, for $G=2$, with 4 rounds, it reaches $10\%$ of the lattice after 522. {\color{black} This indicates the order from the most to the least sensitive to the initial conditions is $G=3$, $G=4$, $G=5$ and $G=2$, which is the same order of the total number of rounds.}

In order to obtain the results depicted in fig. \ref{fig8}, it was necessary to perform more than 1000 simulations, since the simulations ended up becoming identical after a few generation often. It happened in $54.7\%$ of the simulations for $G=2$, $52.1\%$ of the simulations for $G=3$, $53.5\%$ of the simulations for $G=4$ and $59.2\%$ of the simulations for $G=5$.

\begin{figure}[!htb]
	\centering
	\includegraphics{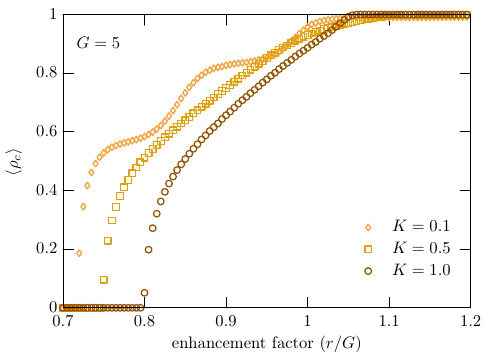}
	\caption{Average density of cooperators for over $5000$ simulations after $5000$ generations as a function of $r/G$ for $G=5$ parametrized by the irrationality coefficient $K$. Here we have set $L=500$.}
	\label{fig9}
\end{figure}

\begin{figure}[!htb]
	\centering
	\includegraphics{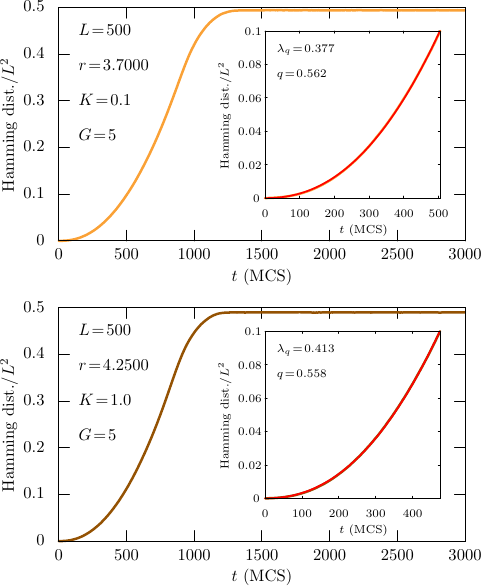}
	\caption{Hamming distance per $L^2$ for $G=5$ as a function of the time. These pictures were obtained by averaging the Hamming distance for over 500 simulations. The inserts show the fits of the $q$-exponential for the first $500$ generations. The parameters used in these simulations are also shown in the figure.}
	\label{fig10}
\end{figure}

{\color{black} The irrationality coefficient, $K$, also plays an important role when it comes to the chaotic behavior of PGG model. In order to visualize its effect over the chaotic behavior we performed simulations for $G=5$ varying the parameter $K$. As it is suggested in fig. \ref{fig9}, the irrationality prevents the systems to cooperate the higher it is. 

Since in fig. \ref{fig9} we used $K=0.1, 0.5$ and $1.0$, in fig. \ref{fig10} we also displayed the Hamming distance for $G=5$ as a function of the time  for $K=0.1$ and $K=1.0$. The inserts in the this picture show the parameters obtained by fitting the Hamming distance for over $1000$ simulations. As it can be noticed, the Hamming distance reaches $10\%$ of the lattice for $K=1.0$ after $475$ generations and for $K=0.1$ it takes $505$ generations to reaches the same $10\%$ of the lattice.}

\section{Conclusion}

In this work, we have studied the chaotic behavior of public goods game, investigating the time evolution of the Hamming distance parameterized by the group size $G$. The main results are displayed in fig. \ref{fig8}. Interestingly, the profile of the Hamming distance is similar to the cases found before in Refs. \cite{2017-Bazeia-SR-7-44900,2017-Bazeia-EPL-119-58003}, once again indicating universality to the Hamming distance behavior. In the present work, however, we have also studied the Lyapunov coefficient, exactly fitting the curve of the Hamming distance with the $q$-exponential function usually found in the case of the Tsallis statistics \cite{Tsallis1988}. The main results suggest that the public goods game here considered engenders weakly-chaotic behavior, {\color{black}with the spreading of small perturbations evolving as faster as one increases the total number of round each player participates in.} 

{\color{black} The study of systems defined on lattices with distinct topologies seems to be interesting possibility of continuation of the present investigation. We can think on a three-dimensional cubic lattice sized $L\times L\times L$ with an arrangement with six neighbours, considering distinct values for $G$, the size of the group. Other collective or social games such as the prisoner’s dilemma game Ref. \cite{rapoport1965prisoner} is also possible to be examined under similar questioning. In this case, the recent study \cite{HAN2023113892} which adds asymmetric players to choose strategies in the conflict, is another possibility. These and other similar systems are presently under consideration, and we hope to report on them in the near future. }

\acknowledgments 
This work is supported by Conselho Nacional de Desenvolvimento Cient\'\i fico e Tecnol\'ogico (CNPq, Grants 303469/2019-6 (DB), 309835/2022-4 (BFO) and 304544/2019-1 (BFO)), Funda\c c\~ao Arauc\'aria, Funda\c c\~ao de Apoio a Pesquisa da Para\'\i ba (FAPESQ-PB, Grant 0015/2019) and INCT-FCx (CNPq/FAPESP).

\end{document}